\documentstyle[11pt]{article}
\begin{document}
\date{}

\title{New integrable string-like fields in 1+1 dimensions}

\author{D.K.Demskoi${}^1$, A.G.Meshkov${}^1$ \\
${}^1${\it Oryol State University, Russia}}
\maketitle
\newfont{\Bbb}{msbm10 scaled1200}
\newfont{\efb}{eufb10 scaled1200}
\begin{abstract}
The symmetry classification method is applied to the
string-like scalar fields in two-dimensional space-time.
When the configurational space is three-dimensional and reducible we present
the complete list of the systems admiting higher polynomial symmetries of
the 3rd, 4th and 5th-order.
\end{abstract}
\section{Introduction}
It is known that many integrable by the inverse scattering method nonlinear
systems  possess infinitely many higher symmetries. And vice versa, if a
nonlinear system admits higher symmetries then it is integrable as a rule.
Therefore, search of the higher symmetries is one of the  methods for the
classification of nonlinear integrable systems. In this report we deal with
the higher local symmetries for the string-like two-dimensional scalar fields:
\begin{equation}
L=(1/2)g_{\alpha\beta}(u)u^\alpha_x u^\beta_t +f(u).         \label{eq:1}
\end{equation}
Here\ $f$\ and\ $g_{\alpha\beta} = g_{\beta\alpha}$\ are some differentiable
functions, $\alpha=1,\dots,m$, and\ $\det(g_{\alpha\beta})\neq0$. Summation
over the repeated indices is implied throughout the article. The field
equations take the following form
\begin{equation}
u^\alpha_{xt}+\Gamma^\alpha_{\nu\mu}(u)u^\nu_x u^\mu_t=f^\alpha(u),\label{eq:2}
\end{equation}
where\ $\Gamma^\alpha_{\nu\mu}$\ are the Christoffel symbols for the metric
\ $g_{\alpha\beta},\ f^\alpha=g^{\alpha\beta}f_\beta,\
f_\alpha=\partial_\alpha f=\partial f/\partial u^\alpha$. We assume that the
Riemann tensor
$$
R^\alpha_{\nu\beta\mu}=\partial_\beta\Gamma^\alpha_{\nu\mu}-
\partial_\mu\Gamma^\alpha_{\nu\beta}+\Gamma^\alpha_{\beta\lambda}
\Gamma^\lambda_{\mu\nu}-\Gamma^\alpha_{\mu\lambda}\Gamma^\lambda_{\beta\nu}
$$
of the configurational space {\Bbb V}$_m$ does not vanish. The higher
symmetries and higher conservation laws for the the systems (\ref{eq:2})
were studied in \cite{Get} and \cite{M}. In the first article the polynomial
conserved densities of the system (\ref{eq:2}) in {\Bbb V}$_2$ were calculated
and all corresponding systems (\ref{eq:2}) were found. In the article \cite{M}
the structure of the polynomial higher symmetries of the system (\ref{eq:2})
in {\Bbb V}$_m$ was investigated, and it was proved that the systems
(\ref{eq:2}) in {\Bbb V}$_2$ admiting the polynomial higher symmetries are the
same as in \cite{Get}. Moreover some preliminary results were obtained for
the reducible space {\Bbb V}$_3$ ($u^\alpha=\{u,v,w\}$):
\begin{equation}
d s^2=d u^2+ 2 \psi(v, w) dv dw                                 \label{eq:3}
\end{equation}
Here we present the systems (\ref{eq:2}) in the space (\ref{eq:3}) admiting
Lie-B\"acklund symmetries of the 3rd, 4th and 5th-order.
\section{General consideration}
Writing out the defining equation for the Lie-B\"acklund symmetries (see
\cite{Ibr}, for example), one can rewrite it in the following matrix form:
\begin{equation}
  (VW + \Phi)\sigma =0.                                          \label{eq:4}
\end{equation}
Here\ $\sigma$\ is the Lie--B\"acklund vector field (the symmetry),
$\Phi^\alpha_\beta=R^\alpha_{\nu\beta\mu}u^\mu_1 u^\nu_2-f^\alpha_{;\beta}$,
and
\begin{equation}
V^\alpha_\beta=\delta^\alpha_\beta D_x +\Gamma^\alpha_{\beta\nu}u^\nu_x,\quad
W^\alpha_\beta=\delta^\alpha_\beta D_t
+\Gamma^\alpha_{\beta\nu}u^\nu_t.                                \label{eq:5}
\end{equation}
The semicolon denotes the covariant differentiation on\ {\Bbb V}$_m$. Symbols
$D_x$ and $D_t$ are the total differentiation operators, for example,
$$
D_x F(x,\,u,\,u_x,\,u_t)=\frac{\partial F}{\partial x}+\frac{\partial F}
{\partial u^\alpha}u^\alpha_x+\frac{\partial F}
{\partial u^\alpha_x}u^\alpha_{xx}+\frac{\partial F}
{\partial u^\alpha_t}(f^\alpha-\Gamma^\alpha_{\nu\mu}u^\nu_x u^\mu_t).
$$

Let us consider the manifold {\efb M} on the jet space corresponding to the
system (\ref{eq:2}). It is obvious that quantities $t,\,x,\,u^{\alpha}_i=\partial_x^i
u^\alpha$ and $\tilde u^\alpha_j=\partial_t^j u^\alpha$ are independent
variables on {\efb M}, and quantities $u^\alpha_{i,j}=\partial_x^i
\partial_t^j u^\alpha$ are dependent.

One of the authors proved earlier \cite{M} the following statements:\\[1mm]
{\bf Theorem 1} {\it Any symmetry $\sigma$\ of the system (\ref{eq:2})
independing on $x, t$\ takes the following form:
$$
\sigma(u,\tilde u)=\tau(u)+\omega(\tilde u),
$$
where the vector fields\ $\tau$\ and\ $\omega$\ satisfy the following
equations:
$$
A_\nu(u)\tau^\alpha+R^\alpha_{\beta\gamma\nu}u^\beta_1\tau^\gamma
=p^\alpha_\nu(u_0)-h^\alpha_{\beta\nu}(u_0)u^\beta_1,\quad
B(u)\tau^\alpha-f^\alpha_{;\gamma}\tau^\gamma=q^\alpha_\nu(u_0)
u^\nu_1-g^\alpha (u_0),
$$$$
A_\nu(\tilde u)\omega^\alpha+R^\alpha_{\beta\gamma\nu}\tilde u^\beta_1
\omega^\gamma=h^\alpha_{\nu\beta}(u_0)\tilde
u^\beta_1-q^\alpha_\nu(u_0),\quad
B(\tilde u)\omega^\alpha-f^\alpha_{;\gamma}\omega^\gamma=g^\alpha (u_0)-
p^\alpha_\nu(u_0)\tilde u^\nu_1,
$$
where $A_\nu$  and $B$ are some linear differential operators.}\\[1mm]
{\bf Theorem 2} {\it  If $\sigma(u,\tilde u)=\tau(u)+\omega(\tilde u)$ is a polynomial of
$u_i$ and $\tilde u_i$, then $\tau(u)$ and $\omega(\tilde u)$ are independent
symmetries.}

 Therefore we may investigate $\tau(u)$ only.\\[1mm]
{\bf Theorem 3} {\it  If the system (\ref{eq:2}) admits the polynomial symmetry $\tau(u)$, then it
admits the homogeneous polynomial symmetries:
\begin{equation}
\tau(u)=a^\alpha_\beta u^\beta_n+A^\alpha_{\beta\gamma}u^\beta_{n-1}
u^\gamma_1+(B^\alpha_{\beta\gamma}u^\gamma_2+C^\alpha_{\beta\nu\gamma}
u^\nu_1 u^\gamma_1)u^\beta_{n-2}+\dots ,                       \label{eq:6}
\end{equation}
where the coefficients depend on the variables $u^\nu$ only.}\\[1mm]
{\bf Theorem 4} {\it  If the symmetry (\ref{eq:6}) is admited, then the
following system
\begin{equation}
a^\alpha_{\beta;\gamma}=0,\quad
X^\alpha_{\beta\gamma;\nu}=a^\alpha_\sigma R^\sigma_{\beta\gamma\nu},\quad
Y^\alpha_{\beta\gamma;\nu}=a^\sigma_\beta R^\alpha_{\gamma\nu\sigma},
                                          \label{eq:7}
\end{equation}
must be solvable.}\\[1mm]
Here the tensors $X$ and $Y$ depend on the coefficients $A$ and $B$ from
(\ref{eq:6}). We have not the general solution of the system (\ref{eq:7}) for
any {\Bbb V}$_m$. But the case of {\Bbb V}$_2$ is investigated
completely.\\[1mm]
{\bf Theorem 5} {\it There are two and only two spaces {\Bbb V}$_2$ where
the system (\ref{eq:7}) has the nontrivial solution. They have the following
metrics
\begin{equation}
(a)\quad d s^2=2 \frac{dv dw}{vw+c},\quad
(b)\quad d s^2=2 \frac{dv dw}{v+w},                     \label{eq:8}
\end{equation}
where $c\ne0$ is a constant.}\\[1mm]
{\bf Theorem 6} {\it  System (\ref{eq:7}) has the nontrivial solution in the reducible space
{\Bbb V}$_3$ if and only if the two-dimensional part of the metric (\ref{eq:3})
is one of the metric (\ref{eq:8}).}
\section{Systems admiting higher symmetries}
In accordance with the results of the section 2 we are going to consider
the field models with the following Lagrangian
\begin{equation}
L=1/2\left[ u_t u_x+ \psi(v,w)(v_t w_x+v_x w_t)\right]+f(u,v,w),\label{eq:9}
\end{equation}
where $\psi=(v w+c)^{-1}$ or $\psi=(v+w)^{-1}$. The field equations take
the following form:
\begin{equation}
u_{tx}=f_u,\ \ v_{tx}=(f_w-\psi_v v_t v_x)/\psi,\ \
w_{tx}=(f_v-\psi_w w_t w_x)/\psi,                           \label{eq:10}
\end{equation}
where the subscripts of $f$ and $\psi$ denote the partial derivatives.

Let us mention from the first, that any system (\ref{eq:10}) where $\psi$ is
arbitrary function, $f=g(v,w)\exp(k u),\,k=const$, and $g$ is arbitrary
function, admits the following higher symmertry:
$$
\sigma^u=(D_x+k u_1) F,\ \ \sigma^v=k v_1 F,\ \ \sigma^w=k w_1 F,\ \
D_t F=0.
$$
The equation $D_t F=0$ is the unique constraint for $F$. It admits
infinitely many solutions and the simplest integral takes the following form:
$$
F=u_2-k \psi v_1 w_1-(k/2) u_1^2.
$$
It is obvious that arbitrary function $\Phi(x,F, D_x F, D_x^2 F,\dots)$
is the integral too. But there exist another integrals with higher order
that are not expressed by the above formula. All these symmetries do not
lead to the integrability and we shall not consider them below. Moreover
we did not consider the case $f(u,v,w)=f_1(u)+f_2(v,w)$ as the independent
equation appear in the system (\ref{eq:10}). Calculations of the higher symmetries
are very cumbersome therefore we used the computer, and we did not calculate
the symmetries with the order more than 5.

There are no the systems (\ref{eq:10}) admiting the 2nd-order symmetry.
If $\psi=(v+w)^{-1}$ then the system (\ref{eq:10}) does not also admit
nontrivial symmetries of the 3rd, 4th or 5th order.

If $\psi=(v w+c)^{-1}$, then the following general form of the 3rd-order
symmetry follows from the equation (\ref{eq:4}):
\begin{eqnarray}
&\sigma^u =& a_1 u_3+(v_2 w_1 c_2+c_3 w_2 v_1) \psi+c_6 u_1^3+
c_7 \psi u_1 v_1 w_1-\psi^2 c_3 w v_1^2 w_1-\psi^2 c_2 v v_1
w_1^2,\nonumber\\[1mm]
&\sigma^v =& a_2 v_3+u_2 c_4 v_1+v_2 (c_5 u_1-3 a_2 v \psi w_1)+
u_1^2 c_8 v_1\nonumber\\[1mm]
&&-2 c_5 v \psi u_1 v_1 w_1+ c_9 \psi v_1^2 w_1+3 a_2 v^2 \psi^2 v_1
w_1^2,\nonumber\\[1mm]
&\sigma^w =& a_2 w_3+u_2 (c_4-c_5) w_1-w_2 (c_5 u_1+3 a_2 w \psi v_1)+
u_1^2 c_8 w_1                                                \nonumber\\[1mm]
&&+2 w c_5 \psi u_1 v_1 w_1+3 w^2 \psi^2 a_2 v_1^2 w_1
+c_9 \psi v_1 w_1^2,                                           \nonumber
\end{eqnarray}
where $a_i$ and $c_i$ are constants. Substituting these functions into the
equation (\ref{eq:4}) we obtained about 60 equations for the function $f$.
Here is the solutions:
\begin{eqnarray}
f&=&   a v \exp(\sqrt{2}u)+b  w \exp(-\sqrt{2}u),             \label{eq:11}\\
f&=& a   v^2 \exp(2u)+ b   w^2 \exp(-2u),                \label{eq:12}  \\
f&=& a   v^2 \exp(2u)+ b   w \exp(-u),                    \label{eq:13} \\
f&=&a v \exp(u)+b w \exp(-u),                                \label{eq:14}\\
f&=& (vw+ {c}/{2})[a \exp(\sqrt{2}u)+b \exp(-\sqrt{2}u)],   \label{eq:15}\\
f&=&a (vw+{c}/{2})\exp(\sqrt{2}u)+b \exp(-\sqrt{2}u),       \label{eq:16}\\
f&=&a (vw+{c}/{2})\exp(\sqrt{2}u)+b \exp(-2 \sqrt{2}u).     \label{eq:17}
\end{eqnarray}

The system (\ref{eq:10}) corresponding to the function (\ref{eq:13}) admits
2-parametric 3th-order symmetry with arbitrary $a_1$ and $a_2$. Other
systems admits unique 3rd-order symmetry, with $a_2\ne0$. But when $a=0$ or
$b=0$, then an additional 3rd-order symmetry appear for each system.
Two systems corresponding to the functions (\ref{eq:14}) and (\ref{eq:16})
admit the 4th-order symmetry.

The system (\ref{eq:10}) corresponding to the function (\ref{eq:11}) with
$b=0$ admits the 7-parametric 5th-order symmetry and does not admit it when
$a b\ne0$. The systems (\ref{eq:10}) corresponding to the functions
(\ref{eq:12}), (\ref{eq:14}) or (\ref{eq:16}) with $b=0$ admit the
4-parametric 5th-order symmetries and do not admit them when $a b\ne0$.
The system (\ref{eq:10}) corresponding to the function (\ref{eq:16}) with
$ab\ne0$ admits the unique 5th-order symmetry. Besides there are only two
new systems admiting the 5th-order symmetries.
The systems (\ref{eq:10}) corresponding to the functions
\begin{equation}
f=a (v^2 w +2 v c/3)\exp(\sqrt{2} u),                          \label{eq:18}
\end{equation}
and
\begin{equation}
f=a v \exp(\sqrt{2/3} u)                                       \label{eq:19}
\end{equation}
admit the 2-parametric 5th-order symmetries.

We believe that the systems corresponding to the functions (\ref{eq:11}) --
(\ref{eq:19}) are integrable by the inverse scattering method and we are going
to find the zero curvature representations for these systems.
\section*{Acknowledgment}
The authors were supported in part by the research grant 96--01--01384
from Russian Foun\-dation for Fundamental Reseach.

\end{document}